\documentclass[preprints,article,accept,moreauthors,pdftex]{Definitions/mdpi} 

\usepackage{threeparttable,booktabs}
  \usepackage{hyperref}
  \hypersetup{
      colorlinks=true,
      linkcolor=blue,
      citecolor=blue,      
      urlcolor=blue
  }

\firstpage{1} 
\pubvolume{xx}
\issuenum{1}
\articlenumber{5}
\pubyear{2019}
\copyrightyear{2019}
\history{Received: date; Accepted: date; Published: date}

\Title{Exploring Basement Surface relationship of north-west Bengal Basin using satellite images and tectonic modeling}

\Author{Sabber Ahamed $^{1, 2}$*, Delwar Hossain $^{3}$ and Jahangir Alam $^{4}$}

\AuthorNames{Firstname Lastname, Firstname Lastname and Firstname Lastname}

\address{%
$^{1}$ \quad Center for earthquake research and information (CERI), The University of Memphis\\
$^{2}$ \quad Now at Asurion, Nashville, Tennessee \\
$^{3}$ \quad Department of Geological Sciences, Jahangirnagar University
Savar, Dhaka, Bangladesh \\
$^{4}$ \quad Geological Survey of Bangladesh, Segunbagicha, Dhaka, Bangladesh}

\corres{Correspondence: sabbers@gmail.com}

\abstract{The Bengal basin is one of the thickest sedimentary basins and is being constantly affected by the collision of the Indian plate with the Burma and Tibetan plates. The northwest part of the basin, our study area, is one of the least explored areas where the shallowest faulted basement is present. Controversies exist about the origin of the basement and its role to the formation of surface landforms. We analyze satellite images, Bouguer anomaly data, and develop a geodynamic model to explore the relationship between the faulted basement and surface landforms. Satellite images and gravity anomalies show a spatial correlation between the surface topography and basement fault structures.  The elevated tracts and the low-lying flood plains are located on top of the gravity highs (horsts) and lows (grabens). The geodynamic model suggests that conjugate thrust faults may exist beneath the horsts that push the horst block upward. Our observations suggest the regional compression and basement faults have a more considerable influence on the development of surface landforms such as the uplifted tracts and the low-lying flood plains.
}

\begin{document}

\section{Introduction}
\label{sec:intro}

\begin{figure}[ht]
\begin{center}
\noindent \includegraphics[scale=0.7]{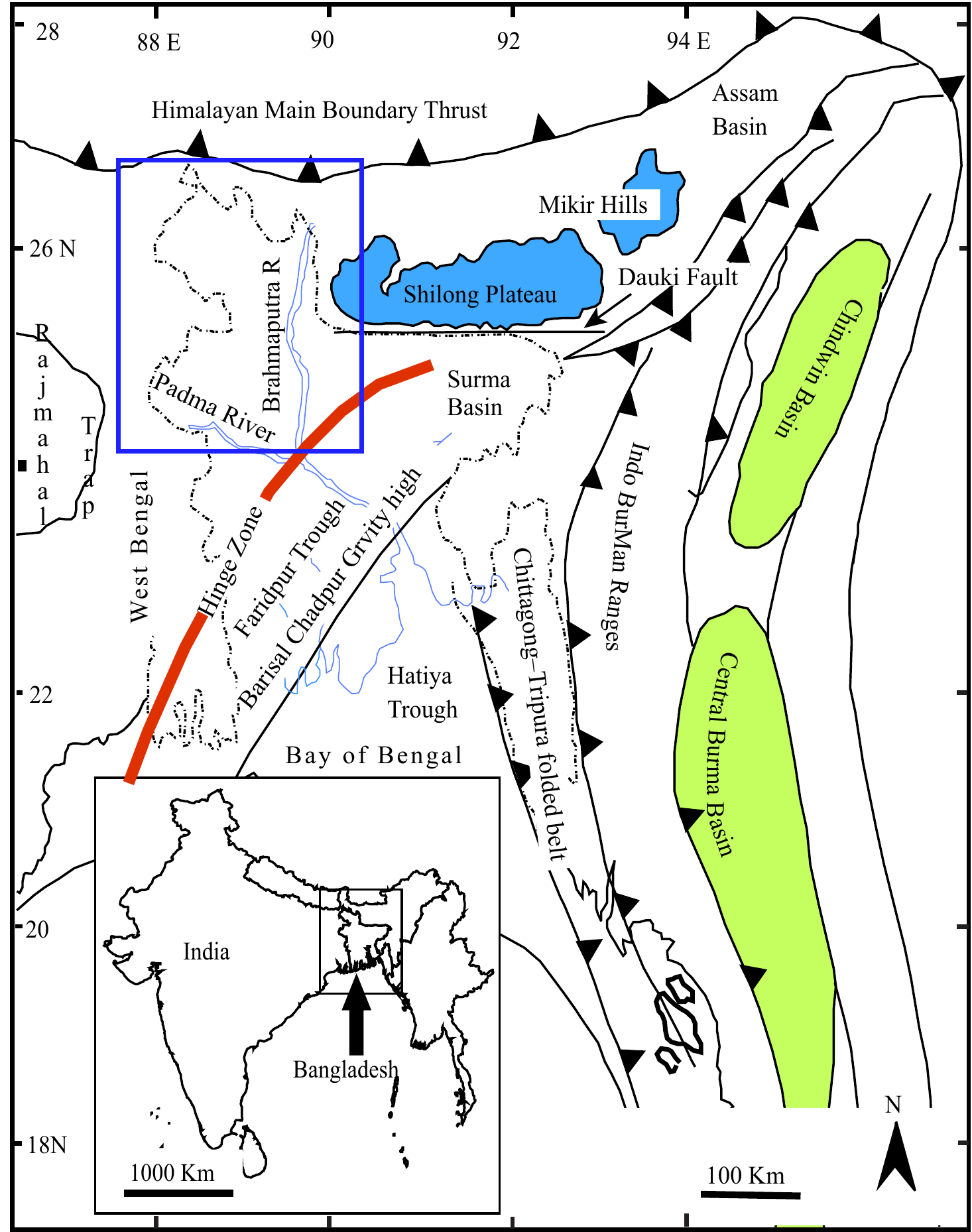}
\end{center}
\caption{\protect\raggedright Tectonic map of Bengal Basin and its surrounding area, modified from~\citet{reimann1993geology, alam1972tectonic, johnson1991sedimentation}. The blue square box is the study area, details shown in Figure~\ref{fig:local_tectonics}. The black line with triangles are the thrust belts. The Shillong plateau is one of the significant structural features and located just right by our study area. The Dauki fault is a major fault along the southern boundary of the Plateau. The thick red line is the Hinge Zone which is an elongated zone separates basin in the east from the shelf zone of the west.}
\label{fig:regional_tectonics}
\end{figure}

The Indian plate was a part of the ancient supercontinent of Gondwana. It started breaking up from Gondwana about 176 million years ago and later become a major plate ~\citep{chatterjee2013longest}. The major tectonic elements of the Indian plate started developing with the northward drift of the Indian plate since Cretaceous and its collision with the Eurasian plate by early to middle Eocene~\citep{sikder20032}. The Bengal Basin (Figure.~\ref{fig:regional_tectonics}) is one of the largest sedimentary basins, located in the north-eastern part of the Indian plate. The Indian shield makes its eastern boundary, whereas the compressional Indo-Burman folded belt makes the western boundary. The Shillong plateau marks one of the significant structural features in the northern portion of the basin.

\begin{figure}[ht]
\begin{center}
\noindent \includegraphics[scale=0.50]{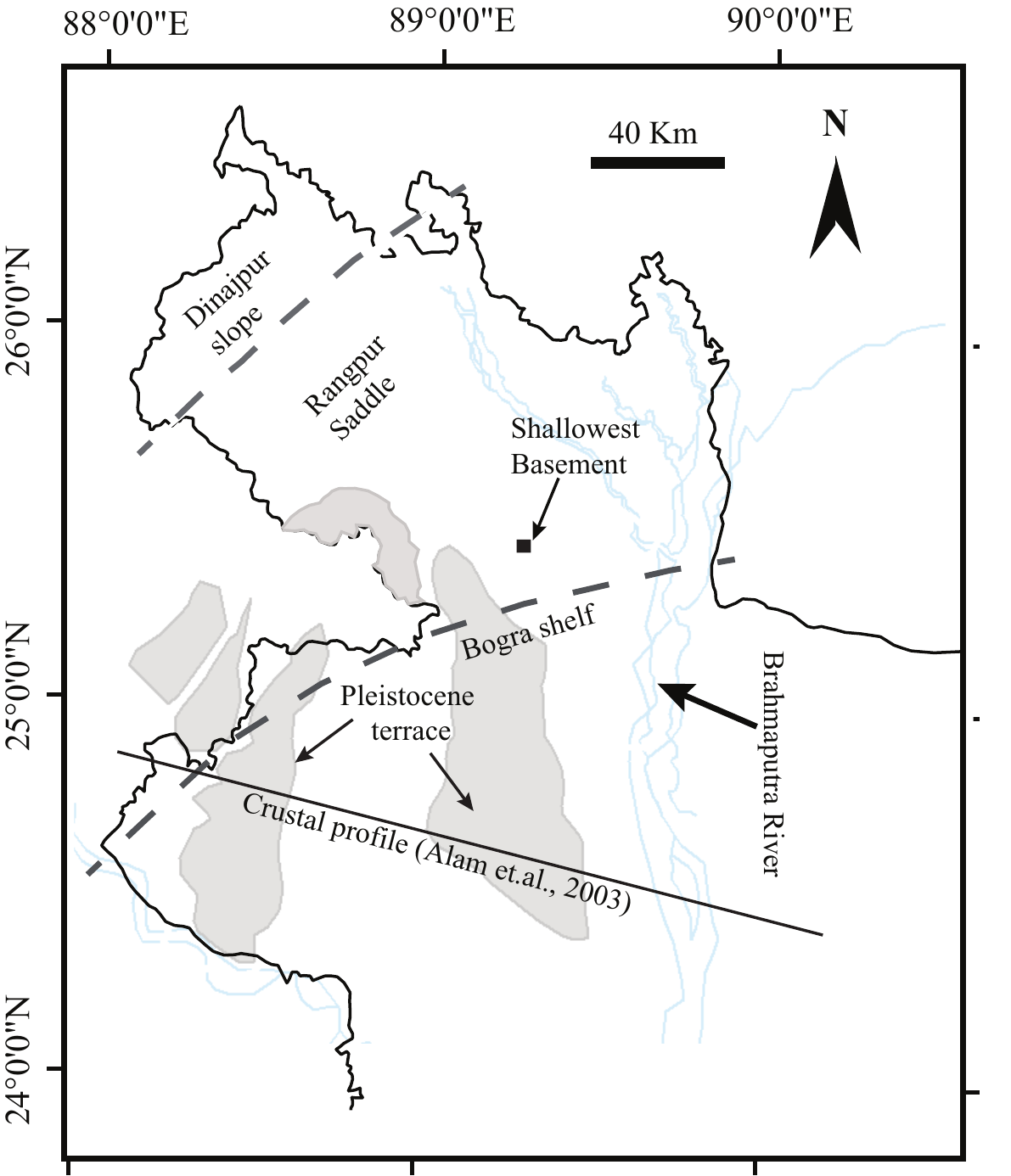}
\end{center}
\caption{\protect\raggedright Major tectonic elements of northwestern Bengal Basin. The region has three distinct geological components (Dinajpur shelf, Rangpur saddle, and Bogra slope) separated by dashed lines. The black square box is the location where the shallowest Paleoproterozoic basement $(\sim128)$m is reported ~\citep{khan1991geology}. A geodynamic model is created to explore the deep basement using the E-W crustal cross-sectional profile used by~\citet{alam2003overview}}
\label{fig:local_tectonics}
\end{figure}

The northwestern part of the Bengal Basin (Figure~\ref{fig:local_tectonics}) has a unique geologic setup. The region has the shallowest Paleoproterozoic basement $(\sim128)$m within the Bengal Basin~\citep{khan1991geology}. Scientists have been researching to understand the origin of the faulted basement. However, no conclusive explanations exist yet. For example, ~\citet{ameen2007paleoproterozoic} and ~\citet{hossain2007palaeoproterozoic} seperately studied the basement rock from Maddhapara, northwestern Bengal Basin to explain the tectonic evolution of the region. Uisng the SHRIMP U–Pb dating technique, both of the studies found a roughly consistent age ($1722\pm6$ Ma, and $1730\pm11$ Ma, respectively). However, their argument on the tectonic evolution of the basement differ from each other significantly.~\citet{ameen2007paleoproterozoic} argued that there is no comparable age found in Indian tectonic zone on the west and Shillong plateau on the east. They proposed that the basement is a discrete trapped micro-continental block. Contrarily, ~\citet{hossain2007palaeoproterozoic} concluded that similar ages are found in the rocks from the Indian tectonic zone and Shillong Plateau. They concluded that the basement in the northwest Bengal basin is the continuation of the greater Indian tectonic zone. We believe that both of the hypothesizes are based on missing decisive data or observations. Therefore more data and studies are required to make a conclusive decision about the tectonic evolution of the northwestern part of the basin.

Another disagreement of this region comes on the origin of the elevated Pleistocene terrace  (Figure~\ref{fig:local_tectonics}). The terrace is locally known as the Barind tract. There are also two groups of opinion exists on the origin of the tract. One group believes that the tract was created by the combined effect of the basement faults and regional compressio. Using aerial photographic interpretations~\citep{morgan1959} showed that Quaternary tectonic activities are responsible for the formation of the tract. Supporting the idea of~\citep{morgan1959},~\citet{hussain2001geological} also concluded that the tract is the product of vertical movements. Recently~\citet{Rashid2015a} studied the region using borehole data and developed stratigraphy of the region. The authors found an abrupt and unusual variation in thicknesses of sedimentary covers that are correlated with the basement structures. The other hypothesis comes from~\citet{monsur1995introduction}. The author argued that the tract has no connection with the regional compression; instead, it is likely to be an erosional geomorphic feature.

In this paper, we explain the interaction between basement faults and surface landforms using different types of data and approaches. We use satellite images to learn about the geomorphic process of the tract and Bouger gravity anomaly data to study the shallow and deep basement structures. Finally, we develop a geodynamic model to explain the dynamic relationship between the basement and surface of the tract. %

\section{Geology and tectonics of the study area}
The northwest part of the basin, the study area, locally known as stable-platform and have three geological components: Dinajpur shelf, Rangpur saddle, and Bogra shelf. The basement in the Dinajpur Shelf gently plunges northward with Himalayan Foredeep, which is approximately 1-3 degrees~\citep{hossain2019synthesis} and is covered by recent sedimentary deposits~\citep{reimann1993geology}. The Rangpur saddle, the southern block of the Dinajpur shelf, connects the Indian Shield and the Shillong Plateau~\citep{hossain2018petrology}. The area has the shallowest basement in the Bengal Basin. The southern slope of the Rangpur Saddle, the Bogra shelf, has numerous graben, half-graben, and horsts. These faults were formed during the rifting process of the Indian plate from Gondwana in the Early Cretaceous~\citep{alam2003overview,reimann1993geology}. 

Two other important geomorphological elements in the area are elevated Pleistocene Barind tract and the Brahmaputra river (Figure.~\ref{fig:local_tectonics}). The tract is the triangular wedge of landmass formed during the Pleistocene. The surface is composed of loose sediments~\citep{Rashid2015a, Rashid2006a}. The Brahmaputra river runs parallel to the eastern side of the tract and is believed to be linked with the lithospheric flexure of the underlying basement ~\citep{rajasekhar2008crustal}.

\section{Data and pre-processing}
We use Landsat thematic mapper satellite images of four different years (1972, 1989, 2003, and 2010) to investigate the temporal surface processes in the study area (Figure~\ref{fig:satellite_images}). Our visual interpretation relied on pixel values of the images and their relationship to local geologic features. For example, water bodies are generally dark in the images, and their corresponding pixel values are around zero. On the other hand, low moisture content features appear as gray to white, and their corresponding pixel values are roughly around 255. We used the Universal Transverse Mercator (UTM) projection system in all the satellite images and gravity maps.

The Bouguer anomaly data published by ~\citet{rahman1990bouguer} are used to study the basement of the area.  The digital contour map of the gravity data was collected from the website of the Department of Interior, USA (\url{https://catalog.data.gov/dataset/bouguer-gravity-anomaly-map-of-bangladesh-grav8bg}). The contour map was converted to point data. Later, for convenience, we converted point data to an equally spaced grids using the kriging interpolation method. All the analysis was performed on the gridded data.
\section{Results and discussion}
In this section, we discuss the surface and subsurface structures and threir relationship. We first use time-series satellite images to explore the time-dependent geomorphic processes. Then we use gravity anomaly data for the shallow and deep basement structures analysis. Finally, we develop a geodynamic model to learn about the dynamic relationship between faulted basement with surface topography.
\subsection{Geomorphic process}
\begin{figure}[ht]
\begin{center}
\noindent \includegraphics[scale=0.60]{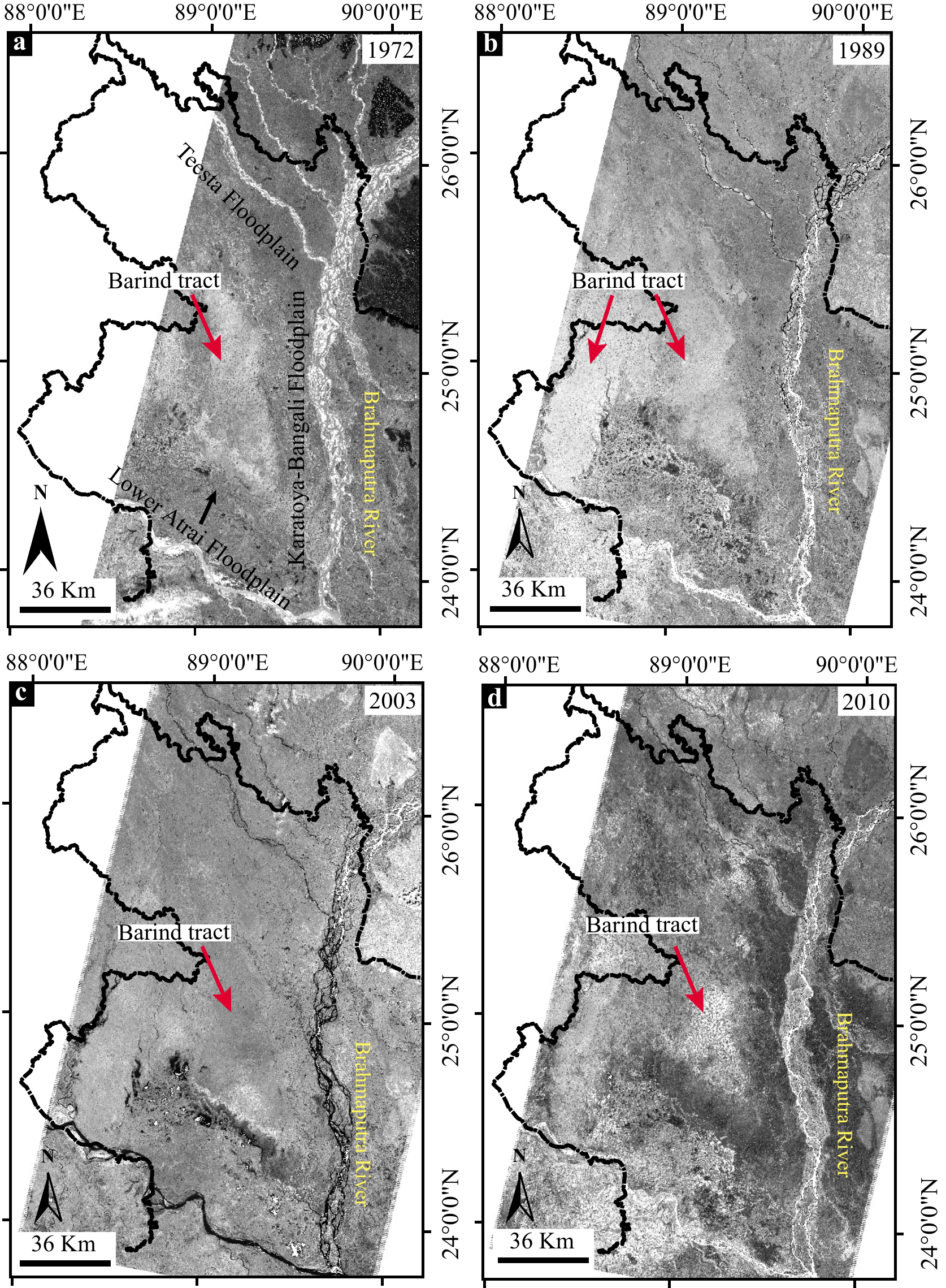}
\end{center}
\caption{\protect\raggedright Time-series Landsat satellite images of the study area. Floodplains are named after~\citet{brammer1996geography}. Red arrows show the location of Barind tract. This figure and associated running/plotting scripts available under~\citet{Ahamed2017}.}
\label{fig:satellite_images}
\end{figure}
Figure.~\ref{fig:satellite_images} shows the Landsat satellite images of the region in four different years. The elevated tract is visible in all of the images. The spatial color differences are mostly limited to the tract and its surrounding low lying floodplains. The southern and eastern boundaries have prominent linear sharp color contrasts. A persistent white tone is present throughout the tract, while the surrounding flood plains have a darker tone. ~\citet{alam1995neotectonic} relates this color variation with the amount of moisture content present in the sediments. The region with higher moisture content generally has a darker tone (e.g., floodplains), whereas the low moisture content regions have a white tone (e.g., tract). 

The color contrast between the elevated tract and surrounding low-lying flood plains also provide information about the local geomorphic and tectonic processes.~\citet{rashid2018structure} mapped the contrast areas as a series of lineaments. Surprisingly, the orientation of the lineaments is consistent with the direction of regional N-S and SE-SW stresses. \citet{khandoker1987origin} concluded that the tract is a horst block along with crustal weakness with compensatory subsidence of the bordering regions. We find that these sharp linear boundaries also coincide with the regional and residual gravity anomalies (Figure.~\ref{fig:regional_gravity} and ~\ref{fig:svd_gravity}). 

\subsection{Deep basement structures}
We use the least-square polynomial fitting surface technique to separate deep crustal structures from shallow ones. The technique has been used for enhancing large scale long-wavelength gravity anomaly, thus regional geologic features~\citep{beltrao1991robust, mickus2003gravity, telford1990applied}. In this technique, a polynomial surface with a certain degree is fitted to the Bouguer anomaly ($g(x_i, y_i)$) data. $x_i$  and  $y_i$ are the locations of the anomalies. A polynomial equation with $n^th$ degree is given as:
\begin{equation}
    f(x_i, y_i) = a_0 + a_{2}x_{i} + a_{3}y_{i} + a_{4}x_{i}^2 + ....... + a_{m}y_{i}^n
\end{equation}
Where n is the degree of the polynomial, m is the total number of the terms of the polynomial, $a_0, a_1..... a_m$ are the coefficients. The error between $f(x_i, y_i)$ and $g(x_i, y_i)$ depends on the several factors: quality of the original data, order used in the polynomial, and the magnitude of the area fitted~\citep{telford1990applied}. Coefficients of the equation can be found by minimizing the least square error with trial and error iterations.

\begin{figure}[ht]
\begin{center}
\noindent \includegraphics[scale=0.60]{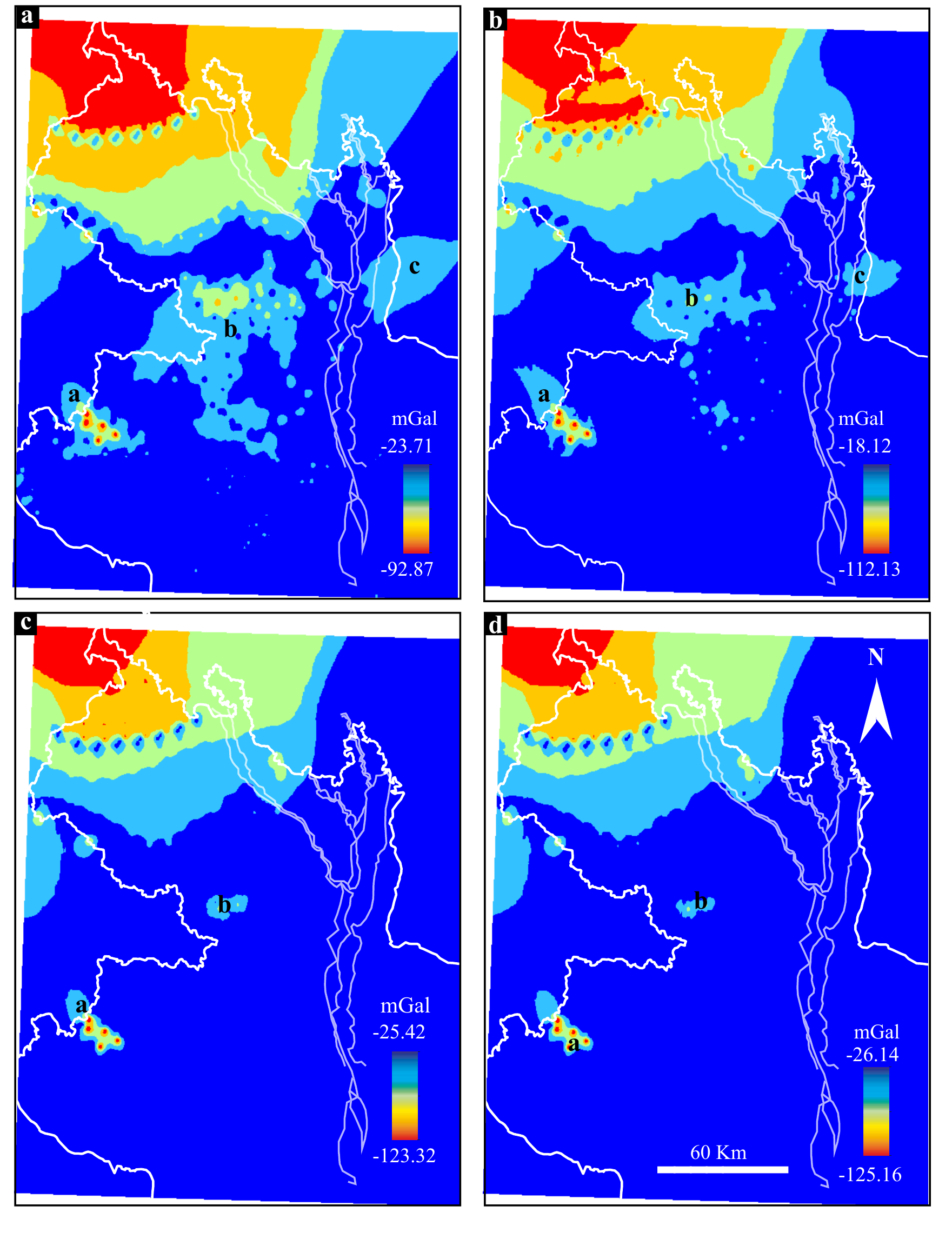}
\end{center}
\caption{\protect\raggedright Regional Gravity anomaly map of the polynomial surface of a) second b) third c) fourth and d) fifth degree. This figure and associated running/plotting scripts available under ~\citet{Ahamed2017}.}
\label{fig:regional_gravity}
\end{figure}

Figure~\ref{fig:regional_gravity} shows the regional gravity anomalies at four different polynomial degrees (second to fifth). All the orders show a strong, negative northeast trending regional gravity anomalies with a non-uniform gradient in the northwestern region. The anomalies have high intensity and long-wavelength. Geologically the region is located on the northern slope of the \textit{Rangpur saddle}. The saddle connects the Shillong Massif and the Mikir hills to the east.~\citet{rahman1990bouguer} shows that these high magnitude anomalies are due to the combined effects of thick low-density sedimentary rocks and a north-dipping, denser substrate in the Himalayan collision zone.
The polynomial fitted gravity data suggest that the region has a deep geologic structure that may extend up to the surface of the area, which is similar to observations from the previous studies~\citep{Rashid2015a, rahman1990bouguer, reimann1993geology}. With the increasing degree of polynomial order, gravity highs disappear from the \textit{Rangpur saddle} except some scattered and moderate magnitude regional gravity anomalies. For interpretation convenient, we denote the gravity highs as a, b, c (Figure.~\ref{fig:regional_gravity}). The highs are also distinguishable on the residual gravity anomaly map (Figure.~\ref{fig:svd_gravity}). Residual gravity anomalies generally represent shallow geologic structures~\citep{telford1990applied}. Indeed, the shallowest basement of the entire Bengal basin has been found in the gravity high $b$, where the reported depth is $\sim128$ m~\citep{ameen2007paleoproterozoic, hossain2007palaeoproterozoic}. From the satellite images (Figure ~\ref{fig:satellite_images}) and regional polynomial surface fitted map, it is obvious that the Barind tract is located on the top of the gravity high a and b. The spatial correlation between the tracts and the gravity highs indicates that the barind tract is connected to the deeper basement structures.

\subsection{Shallow basement structures}
\begin{figure}[ht]
\begin{center}
\noindent \includegraphics[scale=0.60]{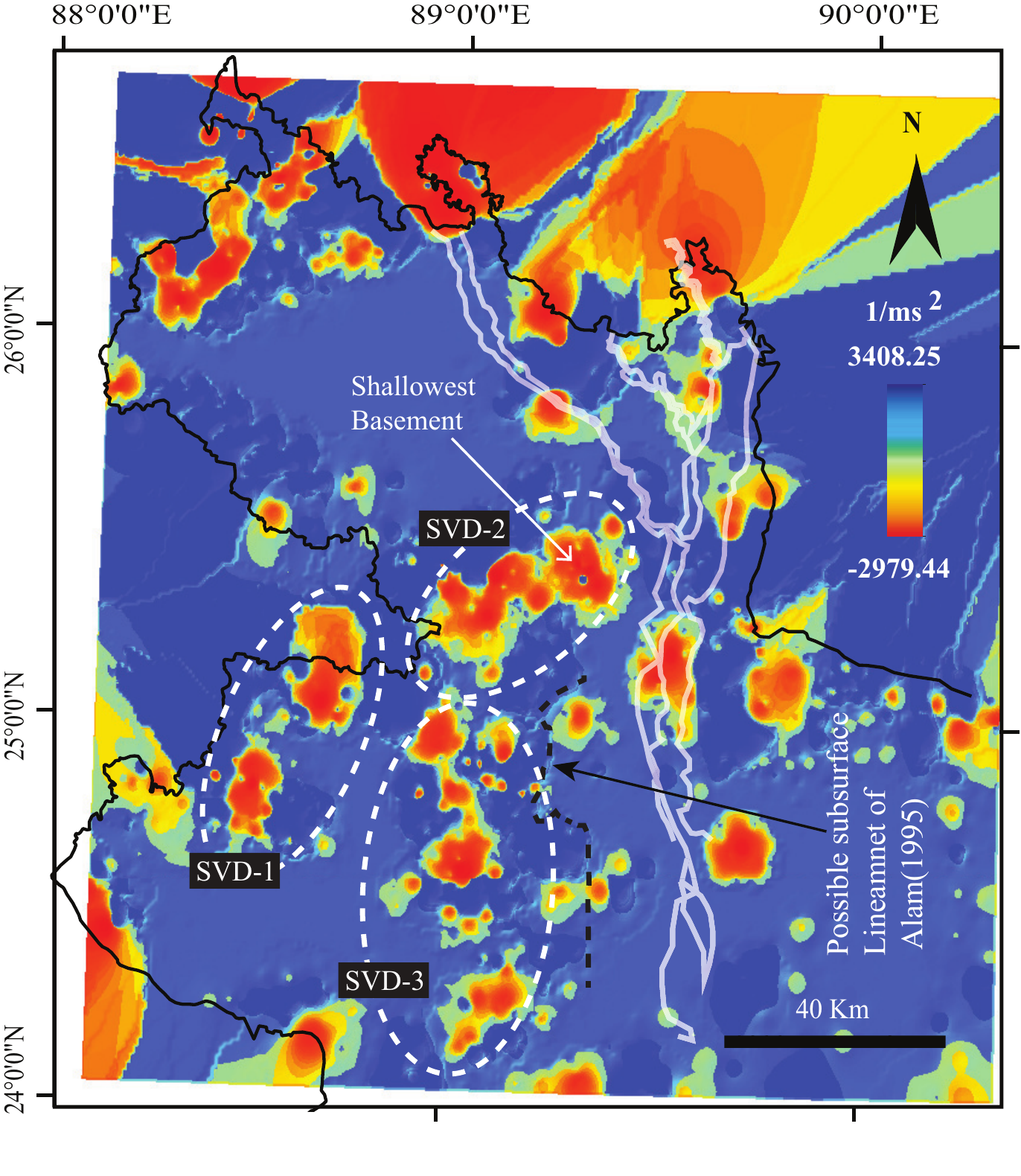}
\end{center}
\caption{\protect\raggedright Second Vertical Derivative (SVD) of Bouguer anomaly showing shallow crustal features. This figure and associated running/plotting scripts available under ~\citet{Ahamed2017}.}
\label{fig:svd_gravity}
\end{figure}

Due to the presence of the shallow basement, we analyzed the residual gravity anomalies that represent the shallow geologic features. The residual anomalies are calculated using Second Vertical Derivative(SVD). The SVD is a measure of curvature, and large curvatures enhance the high-frequency features (near-surface effects) at the expense of deeper or regional anomalies. SVD can be calculated from the second horizontal derivatives~\citep{telford1990applied} as:
\begin{equation}
    SVD = \frac{\partial^2g}{\partial z^2} = -\left(\frac{\partial^2g}{\partial x^2} + \frac{\partial^2g}{\partial y^2}\right)
\end{equation}
Where, $g$ is the Bouguer anomaly at a certain location ($x_i, y_i$). There are many numerical and Fourier transform methods available to compute $\frac{\partial^2g}{\partial z^2}$~\citep{telford1990applied}. In this paper, We used ~\texttt{Oasis montaj} software to calculate the SVD. Figure~\ref{fig:svd_gravity}, shows the SVD of Bouguer anomaly, where several gravity lows surround gravity highs. For interpretation purposes, we group the highs into three clusters of SVDs (SVD-1, SVD-2, SVD-3) (Figure~\ref{fig:svd_gravity}). SVD-1 is located in the northernmost $\textit{Rangpur saddle}$. The location and extent of the SVD-1 are correlated with the regional gravity anomaly-a (Figure ~\ref{fig:regional_gravity}). Again the spatial correlation between regional and residual correlation that the surface landforms are connected to deep graben like depressed structures. Our observation is consistent with the~\citet{reimann1993geology}$'$ hypothesis that the structures may be the N-S aligned grabens with Gondwana fill.

Another group of scattered gravity highs (SVD-2) are present at the southern slope of the saddle, where the shallowest basement in Bengal Basin has been reported~\citep{khan1991geology}. The third group of gravity highs (SVD-3) has a  sharp boundary with gravity lows on the eastern side of the tract. The Brahmaputra river flows through the boundaries of these highs and lows. The river has been moved back and forth many times. Based on the sedimentological studies ~\citet{akter2015evolution, goodbred2003controls} linked the frequent changes of the river course with the local tectonic activities.~\citet{fergusson1863recent, brammer1996geography, allison1998geologic} reported that the recent river diversion happened in 1782 due to an earthquake that occurred on Dauki fault. The authors mentioned that the earthquake created an upward vertical displacement that might have been responsible for the diversion of the river. Our residual gravity anomaly map shows that river flows along with the gravity lows, which are separated by gravity highs on the eastern side of the tract. This suggests that the river flows through the fault zone or depressed graben structures.
\subsection{Basement-surface relationship}
\begin{figure}[ht]
\begin{center}
\noindent \includegraphics[scale=0.7]{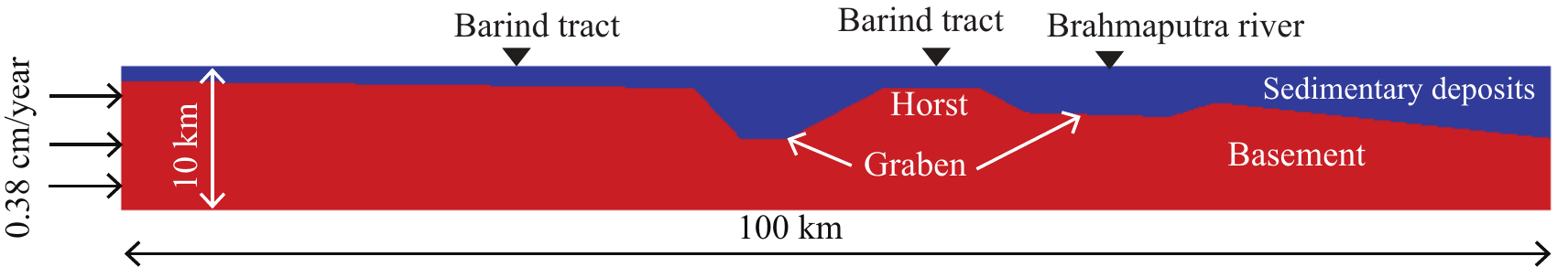}
\end{center}
\caption{Model setup for geodynamic simulation. The model is 100 km long 10 km thick. The model was created using the E-W crustal cross-sectional profile used by~\citet{alam2003overview}. Surface landforms Barind tract and the Brahmaputra river are located on the horsts and grabens. The left boundary is pushed at 0.38 cm/year while the right boundary is kept as a free slip, the bottom boundary is supported by the Winkler foundation \citep[pp.95]{watts2001isostasy}, and the surface is a free surface. This figure and associated running/plotting scripts available under ~\citet{Ahamed2017}.}
\label{fig:faultModelSetup}
\end{figure}

The satellite images and gravity data analysis shows that surface features like the Barind tract, Brahmaputra fault, and subsurface structures horsts and graben spatially correlated. However, it is still not obvious if subsurface geologic structures (horsts and grabens) combined with tectonic activities, can produce the surface features. To test the hypothesis, we created a geodynamic model. The model is initially 100 km long and 10 km thick and has a faulted granitic basement overlain by sediments (Figure~\ref{fig:faultModelSetup}). The model was created using the E-W crustal cross-sectional profile used by~\citet{alam2003overview}. The profile is shown on Figure \ref{fig:local_tectonics}. We solve the energy balance, mass conservation, and momentum balance equations to simulate the model, which is is a Mohr-Coulomb elastoplastic layer.
The energy balance equation~\citep{Ahamed2017} is:
\begin{equation} \label{energy_balance_equation}
(\rho c_p+ 3P\alpha)\frac{dT}{dt}
=
\boldsymbol{\sigma} :\dot{\boldsymbol{\epsilon}}_p
+
3T\alpha\frac{dP}{dt}
-
3PT\frac{\alpha}{\rho}\frac{d \rho}{dt},
\end{equation}
Where $\rho$ is the density, $c_p$ is the specific heat at constant pressure, $P$ is the pressure, $\alpha$ is the volumetric thermal expansion coefficient, $T$ is the temperature, $\boldsymbol{\sigma}$ is the Cauchy stress, $\dot{\boldsymbol{\epsilon}}_p$ is the plastic strain rate tensor and $t$ is the time.
The mass conservation equation~\citep{Ahamed2017} is given as:
\begin{equation} \label{mass_conservation_equation}
\frac{d \rho}{dt} =-\rho \left( \alpha \frac{dT}{dt}+\frac{1}{K}\frac{dp}{dt} \right).
\end{equation}
Where $K$ is the bulk modulus. The momentum balance equation is given as:
\begin{equation} \label{momentum_balance_equation}
    \rho\boldsymbol{\dot{u}} = \nabla\cdot \boldsymbol{\sigma} + \rho g
\end{equation}
$\boldsymbol{u}$ is the velocity vector and  $g$ is the acceleration due to gravity.
Since the profile is at $96^{\circ}$ angle with average Indian plate velocity ($v=3.6 cm/yr$) ~\citep{mahesh2012rigid, socquet2006india}, we use the profile component $(|v\cos(96^{\circ})|$) of the velocity ($v$) to push the left boundary. The right boundary is kept as a free slip. The bottom boundary is supported by the Winkler foundation \citep[pp.95]{watts2001isostasy}, and the surface is free surface. To induce the strain localization, we decrease cohesion to 4 MPa linearly as plastic strain increases to 1. We impose topographic smoothing of the diffusion type with a transport coefficient of $10^{-7} m^2/s$ \citep[pp. 225]{turcotte2014geodynamics}. Parameters used in this model are listed in Table 1:
\begin{table}[ht]
  \label{tab:parametersTable}
  \centering
  \caption{Parameters for the geodynamic model}
  \begin{threeparttable}
  \begin{tabular}{l c c c}
  \hline
  Parameter & Symbol & Sedimentary Layer & Basement\\
  \hline
  Bulk Modulus & $K$ & 7.24 GPa\tnote{a} & 17.89 GPa\tnote{b}\\
  Shear Modulus & $G$ & 1.4 GPa\tnote{a} & 12.5 GPa\tnote{b}\\
  Initial Cohesion & $C$ & 25 MPa & 40 MPa\\
  Friction Angle & $\phi$ & $30^{\circ}$ & $30^{\circ}$\\
  Dilation Angle & $\Psi$ & $0^{\circ}$ & $0^{\circ}$\\
  Density & $\rho$ & 2300 $Kg/m^3$  \tnote{c} & 2750 $Kg/m^3$  \tnote{c}\\
  Volumetric expansion coefficient & $\alpha$ & 3.5 $K^{-1}$ & 3.5 $K^{-1}$\\[1ex]
  \hline
  \end{tabular}

  \begin{tablenotes}
    \item[a] Bulk and Shear modulus of sedimentary are calculated based on density and lower range of P wave$(V_p)$ and $(V_s)$ of porous and saturated sandstone~\citep{bourbie1987acoustics}.\\
    \item[b] Bulk and Shear modulus have been calculated based on Young's modulus$(37583.70)$MPa~\citep{YounusMaddhapara2006} and Poisson's ratio$(0.3)$.\\
    \item[c]~\citet{bourbie1987acoustics}.
  \end{tablenotes}
  \end{threeparttable}
\end{table}

\begin{figure}[ht]
\begin{center}
\noindent \includegraphics[scale=0.7]{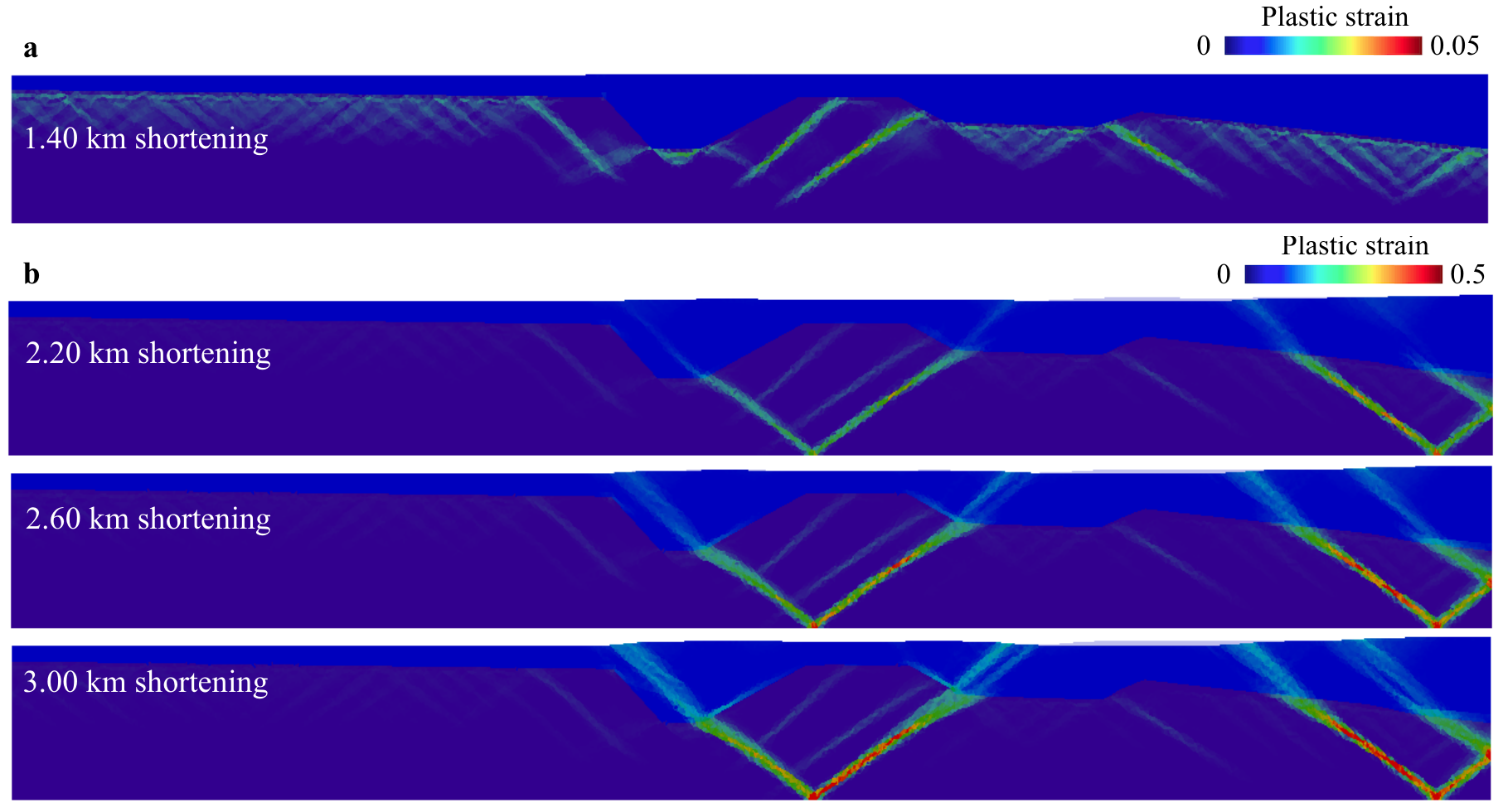}
\end{center}
\caption{Plastic strain distribution of four different shortenings. Platic strain scale a) 0-0.05 and b) 0-0.5. This figure and associated running/plotting scripts available under ~\citet{Ahamed2017}.}
\label{fig:plastic_strain}
\end{figure}

Figure.~\ref{fig:plastic_strain} shows the plains strain distribution at the different shortening of the region. Plastic strain or deformation is the permanent damage to the material. From the beginning, the plastic strain accumulation is concentrated along with the basement and sedimentary deposits interface. Conjugate thrust faults with large and thick plastic strain concentrations start to form only inside the elevated blocks(horsts) of the basement. Comparing to this wide plastic strain concentrated fault, a thin and low amount of plastic strain faults are present subsided regions (grabens) of the basement (Figure.~\ref{fig:plastic_strain}a). At 1.40Km shortening, conjugate thrust faults start to extend deeper. These faults take advantage of the existing horst of the basement.
The model shows that the conjugate thrust faults are responsible for the formation of the surface landforms. Faults first form inside the basement, and with time they reach the surface. The regional compression is also responsible for activating the faults and pushing the horst blocks upward (Figure.~\ref{fig:plastic_strain}b). Since the deformation is accommodated mostly by the conjugate faults in the horst area, the grabens are least affected. That is why we do not see any noticeable upliftment beneath the grabens (Figure.~\ref{fig:plastic_strain}b). Uplifting of the horst and subsidence of grabens are consistent with our gravity and geomorphological observations. Above the horsts, most of the gravity highs and the Barind tract are located. Whereas, the Brahmaputra river flows through the region where the gravity lows and grabens are identified.

Therefore, it is reasonable to conclude that the regional compression and the complex basement faults have a more considerable influence on the formation of surface landforms such as Barind Tract and Brahmaputra river.


\section{Conclusion}
We analyze time-series satellite images, Bouguer gravity anomaly data, and construct a long-term tectonic model. Satellite images reveal significant spatial changes in the uplifted Barind tract and its surrounding low-lying subsidence floodplains. The gravity anomalies show that the basement structure may have a relationship with the surface geomorphology. We find that the uplifted Barind tract is located on top of the gravity highs, whereas low-lying flood plains and faults are on the lows. We construct a tectonic model to explore the relation between the surface and basement structures further. The model produces conjugate thrust faults beneath the gravity highs. The faults reach the surface and push the gravity highs block upward with time. However, no prominent upliftment is seen beneath the grabens. We conclude that the elevated surface tract and its surrounding low-lying floodplains are produced by the regional compression, where the existing basement has a significant role.

\section{References}
\bibliography{basement_surface_interactions}

\end{document}